\documentclass[a4paper, oneside, twocolumn, notitlepage, 10pt]{extarticle_ecoc}
\usepackage{ecoc}
\usepackage{comment}
\usepackage{color}
\usepackage{subfigure}
\usepackage{enumerate}
\usepackage{enumitem}
\usepackage{stfloats}
\usepackage{psfrag}
\usepackage{setspace}
\usepackage{balance}
\usepackage{enumitem}
\usepackage{tikz}
\usepackage{pgfplots,amsfonts}
\usepackage{pgfplotstable}
\usepackage{amsmath,amssymb}
\usepackage{xcolor}
\usepackage{graphicx}
\usepackage{tikz}
\usepackage{pgfplots}
\usepackage{float}
\usepackage{balance}
\usepackage{setspace}
\usepackage{xspace}
\usepackage[normalem]{ulem}

\usepackage{dsfont,empheq}
\usepackage{multicol,multirow}

\usepackage{booktabs, cellspace, hhline}

\usetikzlibrary{shapes}
\usetikzlibrary{spy}
\usetikzlibrary{fit}
\usetikzlibrary{shapes.multipart}
\usetikzlibrary{positioning}

\pgfplotsset{compat=1.14}
\usepackage{wrapfig}

\usepackage[acronym,nomain]{glossaries}

\usepackage{arydshln}
\usepackage{booktabs,makecell,multirow,tabularx}
\usetikzlibrary{calc} 

\definecolor{blue}{rgb}{0.38, 0.51, 0.71} 
\definecolor{darkblue}{RGB}{17, 42, 60} 
\definecolor{red}{RGB}{175, 49, 39} 

\definecolor{orange}{RGB}{217, 156, 55} 
\definecolor{green}{RGB}{144, 169, 84} 
\definecolor{palegreen}{RGB}{197, 184, 104} 

\definecolor{yellow}{RGB}{250, 199, 100} 
\definecolor{brokenwhite}{RGB}{218, 192, 166} 
\definecolor{brokengrey}{rgb}{0.77, 0.76, 0.82} 

\definecolor{darkgreen}{RGB}{0, 100, 0}

\newcommand{\ch}[1]{\textcolor{black}{#1}}

\begin{document}
\selectlanguage{english}    


\title{A General Nonlinear Model for Arbitrary Modulation Formats in the Presence of Inter-Channel Simulated Raman Scattering}%


\author{
    Zhiwei Liang\textsuperscript{(1)},
    Bin Chen\textsuperscript{(1),}*, 
    Jiwei Xu\textsuperscript{(1)}, 
    Yi Lei\textsuperscript{(1)},
    Qingqing Hu\textsuperscript{(1)},
    Fan Zhang\textsuperscript{(2)} and
     Gabriele Liga\textsuperscript{(3)}
}

\maketitle                  


\begin{strip}
    \begin{author_descr}
    
        \textsuperscript{(1)}  School of Computer Science and Information Engineering, Hefei University of Technology, Hefei, China
   
        * \textcolor{blue}{\uline{bin.chen@hfut.edu.cn}}

        \textsuperscript{(2)} School of Electronics, Peking University, Beijing, China

        \textsuperscript{(3)} Department of Electrical Engineering, Eindhoven University of Technology, Eindhoven, The Netherlands
    \end{author_descr}
\end{strip}

\renewcommand\footnotemark{}
\renewcommand\footnoterule{}


\begin{strip}
    \begin{ecoc_abstract}
        The four-dimensional nonlinear model is extended to include the inter-channel stimulated Raman scattering, enabling accurate prediction of dual-polarization four-dimensional modulation formats and probabilistically shaped constellations in high-dispersion regimes. 
        The proposed model is validated via comparisons with the split-step Fourier method and enhanced Gaussian noise model. \textcopyright2025 The Author(s)
    \end{ecoc_abstract}
\end{strip}

\section{Introduction}
With the rapid growth of optical network traffic in recent years, ultra-wideband (UWB) transmission systems have emerged as a highly effective approach to increase the throughput of optical links \cite{10302295,Puttnam:22}. Specifically, the feasibility of UWB transmission from C band to C+L+S+E band has been enabled by the availability of ultra-wideband optical amplifiers, such as 
Bismuth-doped fiber or Raman amplifiers \cite{9795656,10117379}. 
However, as the number of optical bands increases, the effects of fiber nonlinearities, including Kerr and inter-channel stimulated Raman scattering (ISRS) effects, become more significant. Therefore, the analysis of limits on the achievable information rates imposed by fiber nonlinearities is crucial for UWB transmission systems.

For the UWB transmission system, a prominent feature is the power transfer from higher-frequency to lower-frequency carriers, known as ISRS. By extending the Gaussian noise (GN) model to account for the ISRS effect, an integral form expression \cite{8351897} was proposed using a triangular approximation of the Raman gain spectrum. Subsequently, a closed-form approximation \cite{Semrau:17} was developed, which was later significantly generalized to incorporate arbitrary launch power distributions, wavelength-dependent dispersion, and a more accurate description of ISRS \cite{8625492}. 
Another closed-form expression \cite{poggiolini2018Arxiv,zefreh2020Arxiv,Poggiolini2022ECOC,9099626} was derived from the incoherent GN model, augmented with machine-learning factors to enhance the accuracy and incorporate NLI coherence effects. Furthermore, a deep operator network-based waveform-level simulation technique \cite{Zhang:23} was proposed for a C+L band wavelength division multiplexed (WDM) system.

To date, the ISRS GN model \cite{Daniel2019JLT} has been extended to incorporate the impact of modulation formats, thereby providing accurate characterization of the NLI power in the UWB transmission system. 
However, these models are currently incapable of supporting \ch{the modulation formats via constellation shaping} in multi-dimensional (MD) space, like general dual-polarization four-dimensional (DP-4D) or probabilistic shaping (PS) formats. Since MD and PS formats have demonstrated substantial performance gains in both the additive white Gaussian noise channel and the optical fiber channel \cite{Bin2023JLT,Bin2025JSAC}.

The contribution of this paper is twofold. First, we extend the 4D nonlinear model \cite{Liang2024JLT} to include the ISRS effect, which accurately characterizes the intra- and cross-polarization effects of 4D constellations in the UWB system. Secondly, we generalize the ISRS 4D model to PS constellations with finite blocklength by introducing time-window averaged moments, following the approach in \cite{Deng2025OFC}. 
We further evaluate the nonlinear tolerance of three type modulation formats (regular QAM, advanced 4D and PS). Simulation results show that the proposed model accurately predicts both Kerr and ISRS effects, with an NLI power coefficient estimation error below 0.1~dB.

\section{Analytical Formulation of the ISRS 4D Model}
\begin{figure*}
    \centering
    \includegraphics[width=\textwidth]{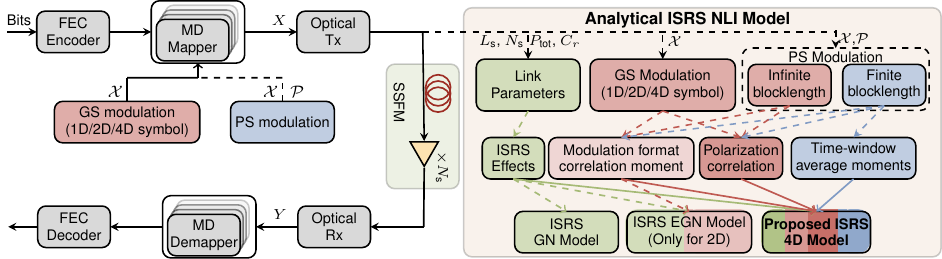}
    \caption{System model under consideration. Left: the block diagram of the considered generic optical fiber system employing various modulation formats. Right: the analytical ISRS NLI model.}
    \label{fig:system}
\end{figure*}

The optical fiber system model under consideration is shown on the left side of Fig.~\ref{fig:system}. At the transmitter, a binary forward error correction (FEC) encoder encodes information bits to generate a binary sequence. This sequence is then mapped into MD symbols $X$ by an MD mapper using the predefined constellation coordinates and probability distribution pairs $\{\mathcal{X},\mathcal{P}\}$ (regular QAM, MD constellation, or PS-QAM). Then the MD symbols are transmitted through the optical fiber channel. At the receiver, the received symbols are processed using ideal chromatic dispersion compensation, matched filtering, sampling, and ideal phase compensation for potential constant phase rotation. The output symbols $Y$ are then processed by a demapper and a binary FEC decoder. 

\ch{The right side of Fig.~\ref{fig:system} depicts analytical models for NLI power coefficient estimation, which replace the time-consuming split-step Fourier method (SSFM) by inputting link parameters and specified modulation formats. The ISRS 4D model accounts for more effects (polarization correlation and temporal symbol correlations) that are neglected in ISRS GN/EGN models, enabling its application to arbitrary modulation formats (including GS and PS constellations).} 
Specifically, the NLI contribution generated on the $i^{\text{th}}$ channel can be found as 
\begin{align}\label{eta}
\begin{split}
     &\eta_i = \frac{8}{9}\frac{\gamma^2}{P_i^3} \Big [\boldsymbol{\Phi}^{\text{SCI}}_{i}(\mathcal{X},\mathcal{P})\chi^{\text{SCI}}(f_i,f_i,N_s,L_s,C_r)\\
    &+ \sum_{k=1,k\neq i}^{N_{ch}}\boldsymbol{\Phi}^{\text{XPM}}_{k}(\mathcal{X},\mathcal{P})\chi^{\text{XPM}}(f_i,f_k,N_s,L_s,C_r) \Big],
\end{split}
\end{align}
where the $N_s$ and $N_{ch}$ represent the number of spans and channels, respectively. The $\gamma$ is the NLI coefficient and $P_i$ is the transmitted power of the $i^{\text{th}}$ channel. The $L_s$ denotes the span length. 
The terms $\boldsymbol{\Phi}^{\text{SCI}}_{i}(\cdot)$ and $\boldsymbol{\Phi}^{\text{XPM}}_{k}(\cdot)$ are several higher-order intra- and cross-polarization moments of modulation formats $\mathcal{X}$, which are given in Table~\uppercase\expandafter{\romannumeral2} \cite{Liang2024JLT}. 
Remarkably, for the short-blocklength shaping schemes, the calculation of $\boldsymbol{\Phi}^{\text{SCI}}_{i}(\cdot)$ and $\boldsymbol{\Phi}^{\text{XPM}}_{k}(\cdot)$ must account for temporal symbol correlations by additionally introducing time-window average moments of PS-4D modulation formats. For more details, we refer the reader to Fig.~2 in \cite{Deng2025OFC}.

The coefficient $\chi^{\text{SCI/XPM}}(\cdot)$ in Eq.~\eqref{eta} is a function of the link parameters. It is important to emphasize that the coefficient $\chi^{\text{SCI/XPM}}(\cdot)$ is  specialized into different forms by a triple of frequencies $(f_1, f_2, f_i)$ and time slot combinations, which can be found in Table~\uppercase\expandafter{\romannumeral3} \cite{Liang2024JLT}. To introduce the ISRS effect, we represent the $\chi(\cdot)$ in a general form, which can be written as
\begin{align}
\begin{split}
    \chi(&f_i,f_k,N_s,L_s,C_r)=\int \int  G_{\text{Tx}}(f_1)G_{\text{Tx}}(f_2)\\
    &\cdot G_{\text{Tx}}(f_1+f_2-f_i)\mu(f_1,f_2,f_i,N_s,L_s,C_r)\\
    &\cdot\mu^*(f'_1,f'_2,f'_i,N_s,L_s,C_r)df_1df_2.
\end{split}
\end{align}

The $\mu(f_1,f_2,f_i,N_s,L_s,C_r)$ denotes four-wave mixing (FWM) efficiency, given by
\begin{align}\label{LinkF}
\begin{split}  
    \mu(f_1,f_2,f_i,&N_s,L_s,C_r) =  \sum_{j=1}^{N_s} \Big [\int_0^{L_s} e^{-j\phi z}  \\
    &\cdot \frac{P_{\text{tot}}e^{-\alpha z-P_{\text{tot}}C_rL_{\text{eff}}(f_1+f_2-f_i)}}{\int G_{\text{Tx}}(v) e^{- P_{\text{tot}} C_r L_{\text{eff}} v dv}} dz\Big],
\end{split}
\end{align}
in which the $P_{\text{tot}}$ is the total transmitted optical power and $C_r$ is the slope of a linear regression of the normalized Raman gain spectrum. The $L_{\text{eff}} = \frac{1-e^{-\alpha z}}{\alpha}$ is the effective length. And the $\phi=4\pi^2(f_1-f_i)(f_2-f_i)[\beta_2+\pi(f_1+f_2)\beta_3]$. The second term is the power profile of the signal along the transmission link. For C-band transmission, the power profile is assumed to decay according to a constant link attenuation factor $\int_0^{L_s} exp(-\alpha z)dz$.

\section{Simulation Results}
In this section, the proposed ISRS 4D model (\tikz{\draw [solid, line width=1.5pt,color = black] (1, 3.5) -- (1.5, 3.5);}) was validated through numerical simulations using SSFM (\tikz{\draw plot[mark=*, mark size=1.5,mark options={color=black, thick}] (1.25, 1);}), and the results were compared with the ISRS enhanced GN (EGN) model (\tikz{\draw [dashed, line width=1.5pt,color = black] (1, 3.5) -- (1.5, 3.5);}). 
Due to the high computational complexity of SSFM for C+L band transmissions, the Raman gain slope $C_r$ was set to 1.12~1/W/km/THz to ensure significant inter-channel power transfer, which is 40 times a more conventional value of 0.028~1/W/km/THz \cite{Hami2020Arxiv}. 
The parameters of the fiber link are as follows: $\alpha$ = 0.2~dB/km, $D$ = 17~ps/nm/km, $\gamma$ = 1.2~(W·km$)^{-1}$. A dual-polarization single-span WDM system with 23 channels was transmitted at a symbol rate of 45~GBaud, a channel spacing of 46~GHz and a root-raised-cosine (RRC) filter roll-off factor of 0.01\%. A 100~km fiber span was followed by an ideal erbium-doped fiber amplifier (EDFA) that fully compensates fiber loss. The launch power was set to 0~dBm per channel. 

The numerical results are obtained for three modulation formats, including PM-8QAM (\tikz{\draw [solid, line width=1.5pt,color = red] (1, 3.5) -- (1.5, 3.5);}), 4D 64-ary polarization ring switched (4D-64PRS) format \cite{Bin2019JLT} (\tikz{\draw [solid, line width=1.5pt,color = blue] (1, 3.5) -- (1.5, 3.5);}) and PS-16QAM which probabilistically shapes PM-16QAM to an entropy of $H(X)$ = 6~bit/4D-sym based on constant-composition distribution matching (CCDM) with blocklength $n=100$ (\tikz{\draw [solid, line width=1.5pt,color = green!50!white] (1, 3.5) -- (1.5, 3.5);}) and $n=1000$ (\tikz{\draw [solid, line width=1.5pt,color = green] (1, 3.5) -- (1.5, 3.5);}). 
The figures show the NLI power coefficient $\eta_i$ in dB$(W^{-2}) = 10\log_{10}(\frac{1}{\text{SNR}_{\text{eff}} P_i^2} \cdot 1W^2)$ as a function of the channel number $i$. The $\text{SNR}_{\text{eff}}$ is the effective signal-to-noise ratio without adding amplified spontaneous emission (ASE) noise.

\begin{figure}
    \centering
    \includegraphics[width=.5\textwidth]{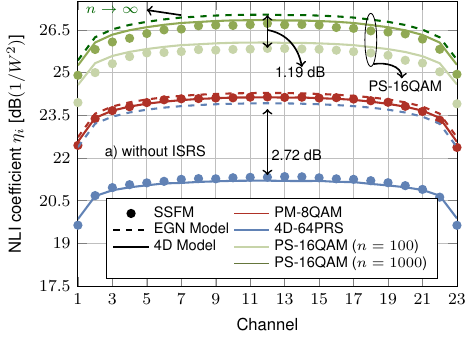}\\
        \includegraphics[width=.5\textwidth]{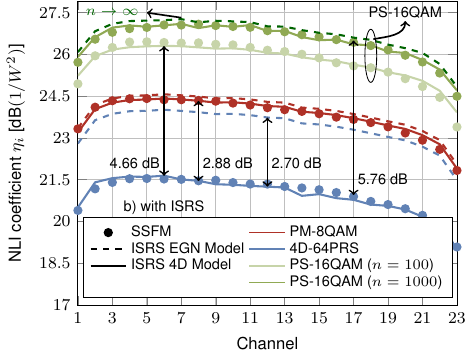}
     \vspace{-1em}
    \caption{The nonlinear interference coefficient $\eta_i$ as a function of the channel index $i$ for three different modulation formats after one span. The results were obtained using three distinct approaches: SSFM, in ISRS EGN model and in the proposed ISRS 4D model using Eq.~(1). Figs. a) and b) show the case without and with considering ISRS, respectively.}
    \label{fig:eta_ch}
\end{figure}

The nonlinear interference coefficients as a function of the channel index without and with ISRS are shown in Figs.~\ref{fig:eta_ch} (a) and (b). 
As shown in Fig.~\ref{fig:eta_ch} (a), the NLI power coefficients are symmetrically distributed from the middle channel for all modulation formats. 
For the PM-2D modulation format, 
\ch{the EGN model and 4D model give the same results as the SSFM.}  
For the DP-4D modulation format, the EGN model leads to inaccuracies of up to 2.72~dB for 4D-64PRS. \ch{This discrepancy originates primarily from the fundamental assumption of statistical independence between the x- and y-polarization states in the EGN model.} 
In addition, since the EGN model operates under the assumption of identically distributed (i.i.d.) input symbols, it is only applicable to PS modulation formats with infinite blocklength ($n\rightarrow \infty$). The results of the EGN model for PS-16QAM with infinite blocklength are shown as the dark green dashed line (\tikz{\draw [dashed, line width=1.5pt,color = darkgreen] (1, 3.5) -- (1.5, 3.5);}).  
However, our proposed 4D model shows great agreement with the numerical results for 4D-64PRS and PS-16QAM with blocklength $n=100$ and $n=1000$. The average mismatch between the ISRS 4D model and SSFM is about 0.1~dB for 4D-64PRS and 0.2~dB for PS-16QAM with blocklength $n=100$ and $n=1000$. 
This can be attributed not only to the 4D model considering polarization correlation. In addition, the 4D model introduces time-window averaged moments of the modulation formats to take the temporal symbol correlations into account. Furthermore, we can observe that the performance of PS-16QAM increases as the blocklength $n$ decreases,  exhibiting a gain of approximately 1.19~dB compared to PS-16QAM with finite blocklength $n=100$. 

As depicted in Fig.~\ref{fig:eta_ch} (b), it can be observed that a frequency-dependent asymmetry is evident in the NLI power coefficient distribution, characterized by higher magnitudes at low frequencies and lower magnitudes at high frequencies. This can be attributed to the ISRS leading to a power transfer from higher to lower frequencies, increasing in strength with wider frequency separation. \ch{The NLI power coefficients are well predicted the results of PM-8QAM by the ISRS EGN model and our proposed model. While the average gap between the ISRS EGN model and SSFM simulations is approximately 2.7~dB for 4D-64PRS.} 
Remarkably, a key observation is the superior nonlinear performance of 4D constellations relative to 2D formats and PS formats, with this advantage persisting across both C-band and UWB transmission regimes. For example, 4D-64PRS demonstrates a 2.88~dB improvement over PM-8QAM, while it shows a 4.66~dB enhancement compared to PS-16QAM with blocklength $n=100$ and 5.76~dB compared to PS-16QAM with blocklength $n=1000$.

\section{Conclusion}
We present a general ISRS 4D model that analytically predicts NLI power with ISRS effects for DP-4D modulation formats and PS constellations in high-dispersion systems. Our numerical results indicate that the proposed model offers good accuracy, as validated by SSFM simulations across three modulation schemes: a conventional 2D constellation (PM-8QAM), a nonlinearity-tolerant 4D modulation format (4D-64PRS) and PS-16QAM with blocklength $n=100$ and $n=1000$, whereas the ISRS EGN model fails to converge. The results also indicated that the 4D constellations have great potential in UWB transmission systems. Therefore, we believe that our model could become a powerful analytical tool for optimizing constellations in 4D space, with the potential to achieve significant NLI mitigation in ultra-wideband systems.

\section{Acknowledgements}
This work is supported by the National Key Research and Development Program of China (No.~2024YFB2908400), the NSFC Program (No.~62171175), the Fundamental Research Funds for the Central Universities under Grant JZ2024HGTG0312, and the State Key Laboratory of Advanced Optical Communication Systems and Networks, China.


\defbibnote{myprenote}{}
\printbibliography[prenote=myprenote]


\end{document}